\documentclass[twocolumn, prl, amssymb,  aps, showpacs,preprintnumbers,
amsmath,showkeys,floatfix]{revtex4}

\usepackage{epstopdf}
\usepackage{graphics}
\usepackage{graphicx}
\usepackage{dcolumn}
\usepackage{bm}
\usepackage{longtable}
\usepackage{epsfig}
\usepackage{times}
\usepackage{url}
\usepackage{color}

\newcommand{\be}{\begin{equation}}
\newcommand{\ee}{  \end{equation}}
\newcommand{\ba}{\begin{eqnarray}}
\newcommand{\ea}{  \end{eqnarray}}
\newcommand{\ve}{\varepsilon}

\begin{document}

\title{Laser-Nucleus Reactions: Population of States far above
Yrast and far from Stability}

\author{A. P\'alffy}
\email{Palffy@mpi-hd.mpg.de}

\author{ H. A. Weidenm{\"u}ller}

\affiliation{Max-Planck-Institut f{\"u}r Kernphysik, 69029 Heidelberg, Germany}

\begin{abstract}
Nuclear reactions induced by a strong zeptosecond laser pulse are
studied theoretically in the quasiadiabatic regime where the photon
absorption rate is comparable to the nuclear equilibration rate. We
find that multiple photon absorption leads to the formation of a
compound nucleus in the so-far unexplored regime of excitation
energies several hundred MeV above the yrast line. At these energies,
further photon absorption is limited by neutron decay and/or induced
nucleon emission. With a laser pulse of $\approx 50$ zs duration,
proton-rich nuclei far off the line of stability are produced.

\end{abstract}




\maketitle

{\it Introduction}. Recent experimental developments in laser physics
hold promise to advance the new field of laser-induced nuclear
reactions beyond so-far explored territory in nuclear physics. Efforts
are under way at both ELI~\cite{Eli12} and IZEST~\cite{Izest} to generate a multi-MeV zeptosecond coherent laser
pulse by backward Compton scattering of optical laser light on a sheet of relativistic
electrons~\cite{Mou11, Mou11a}. 
Although perhaps somewhat ahead of existing technology, this possibility poses a challenge for nuclear
reaction theory which would be confronted with a totally new parameter
regime. The central question is: which reactions are expected to occur
when an intense high-energy coherent laser pulse hits a medium-weight
or heavy target nucleus?  What is the difference between this case and
other areas~\cite{Palffy2008,Wei11,Pia12,Palffy2013} of laser-matter
or laser-nucleus interaction? The answers are relevant also for the
layout of future experiments.

In this Letter we provide first semiquantitative theoretical answers
to these questions by combining a newly developed method of
calculating nuclear level densities at high excitation energies and
for large particle numbers~\cite{Pal13a, Pal13b} with concepts of
nuclear reaction theory. Our work addresses the quasiadiabatic
laser-nucleus interaction regime where excitation and relaxation
processes are governed by similar time scales. We show that coherent
photon absorption by medium-weight and heavy nuclei can produce high
excitation with low angular momentum transfer, leading to compound 
nuclei several  hundred MeV above yrast. Our approach renders
possible the semiquantitative study of the competition between photon
absorption, photon-induced nucleon emission, neutron evaporation, and
fission. The latter turns out not to be competitive. With neutron 
evaporation or photon-induced nucleon emission
overtaking photon absorption at energies below the saturation of the
latter for medium-weight and heavy nuclei, we expect proton-rich
nuclei far from the valley of stability to be produced. Laser-nucleus 
interaction experiments at ELI or IZEST thus promise to shed light on the 
structure of such nuclei and the time scales and level densities involved.

To be specific, we consider a laser pulse containing $N = 10^3 - 10^4$
coherent photons with mean photon energy $E_L \approx 1 - 5$ MeV and
with an energy spread $\sigma \approx 50$ keV (and a corresponding
pulse duration $\hbar / \sigma \approx 10^{- 20}$ s). Nuclei are bound
by the strong interaction. As a consequence, the electromagnetic
interaction of even such a strong laser pulse with a nucleus is much
less violent than the interaction of a medium-intensity optical laser
pulse with an atom. In the atomic case, a laser field strong enough to 
distort the Coulomb potential and set electrons free is characterized by 
an electric field strength roughly given by 
the ratio between the ionization potential and the Bohr radius, i.e.,  $\simeq10^9$ eV/cm. 
For a corresponding distortion of the nuclear
potential, the electric field strength would have to be roughly given by the
ratio between the nucleon binding energy and the nuclear radius, i.e.,
of order $10^{19}$ eV/cm. Despite the MeV photon energy, even the laser pulse 
under consideration here does not produce such strong fields, being
actually rather weak. A quantitative analysis using the Keldysh
parameter~\cite{Kel65} supports this qualitative argument.

For photons in the few-MeV range, the product of wave number $k$
and nuclear radius $R$ obeys $k R \ll 1$. In addition, unlike the case of
low-lying nuclear excitations \cite{Palffy2008}, here available states of all spins
allow the use of the dipole approximation. Four energy scales are relevant for
the laser-nucleus reaction. In addition to the mean laser photon
energy $E_L \approx 1 - 5$ MeV and the energy spread $\sigma \approx
50$ keV, these are the effective dipole width and the nuclear
spreading width. For the effective dipole width of a pulse of coherent photons we use the
semiclassical expression $N \Gamma_{\rm dip}$ valid for $N \gg 1$
coherent photons, with the standard nuclear dipole width $\Gamma_{\rm
  dip}$ in the keV range and $N \Gamma_{\rm dip} \approx 1 - 5$
MeV. In the course of the reaction up to $N_0 \approx 5 \times 10^2$
photons may be absorbed by the nucleus. We neglect the resulting
reduction of $N$ in $N \Gamma_{\rm dip}$. The spreading width
$\Gamma_{\rm sp}$, absent in atoms, accounts for the residual nuclear
interaction. For
excitation energies up to several $10$ MeV, $\Gamma_{\rm sp}$ is of
the order of $5$ MeV~\cite{Har86}.  The nuclear relaxation time 
$\hbar/\Gamma_{\rm sp}$ in which the compound nucleus reaches statistical equilibrium and the mean time for dipole absorption $\hbar/N
\Gamma_{\rm dip}$ are both much shorter than the duration $\hbar /
\sigma$ of the laser-induced nuclear reaction.

The laser-nucleus interaction is characterized by three regimes. (i)
In the perturbative regime $N \Gamma_{\rm dip} \ll \Gamma_{\rm sp}$,
single excitation of the collective dipole mode plays the dominant
role~\cite{Wei11}. The experimental signal for the laser-nucleus
interaction in this regime is the non-exponential decay in time of the
compound nucleus~\cite{Die10}. (ii) In the sudden regime ($N
\Gamma_{\rm dip} \gg \Gamma_{\rm sp}$) the residual interaction is
irrelevant. Nucleons are excited independently of each other and are
emitted from the common average potential. For 
sufficiently long pulse duration, the nucleus evaporates. (iii) The
quasiadiabatic regime ($N \Gamma_{\rm dip} \approx \Gamma_{\rm sp}$)
forms the topic of this Letter and arguably is physically the most
interesting one since it leads to excitation energies far above the
yrast line and to nuclei far beyond the valley of stability. For $N
\Gamma_{\rm dip} \approx \Gamma_{\rm sp}$ nuclear equilibration is as
fast as single-photon absorption. The binding energy of a nucleon $E_b
\approx 8$~MeV being larger than the photon energy $E_L$ considered
here, the dipole excitation energy $E_L$ of the nucleon is shared
almost instantaneously with several or many other nucleons.  This
equilibration mechanism is absent in atoms. Only absorption of a
large number of photons leads to significant induced particle emission
or significant neutron evaporation from the nucleus. Because of these
inherently nuclear processes without atomic counterpart, the
theoretical methods used here to describe the laser-nucleus
interaction do not relate to the strong-field approximation~\cite{SFA}
known from atomic physics. Nuclear photon absorption may rather be
treated in a manner analogous to nucleon-induced precompound
reactions~\cite{Wei08}, i.e., in terms of a (set of) time-dependent
master equation(s). In this Letter we use a simplified version of such
an approach.
 
{\it Quasiadiabatic Regime}. In this regime, the nucleus (almost)
attains statistical equilibrium between two subsequent photon
absorption processes. Consecutive absorption of $N_0 \gg 1$ dipole
photons by an (almost) equilibrated compound nucleus leads to high
excitation energies $N_0 E_L$. Explicit calculation shows that the
average nuclear spin is given by $J = \hbar \sqrt{N_0}$. Therefore,
the laser-induced reactions open access to the regime of states with
small spin far above the yrast line, not accessible to reactions
induced by heavy ions, see Fig.~\ref{1}. Since $J / \hbar$ is only of
order $10$ even for several $100$ absorbed photons (see the inset of
Fig.~\ref{1}), we neglect spin in what follows.

\begin{figure}[b]
\vspace{-0.4cm}

\includegraphics[width=0.5\textwidth]{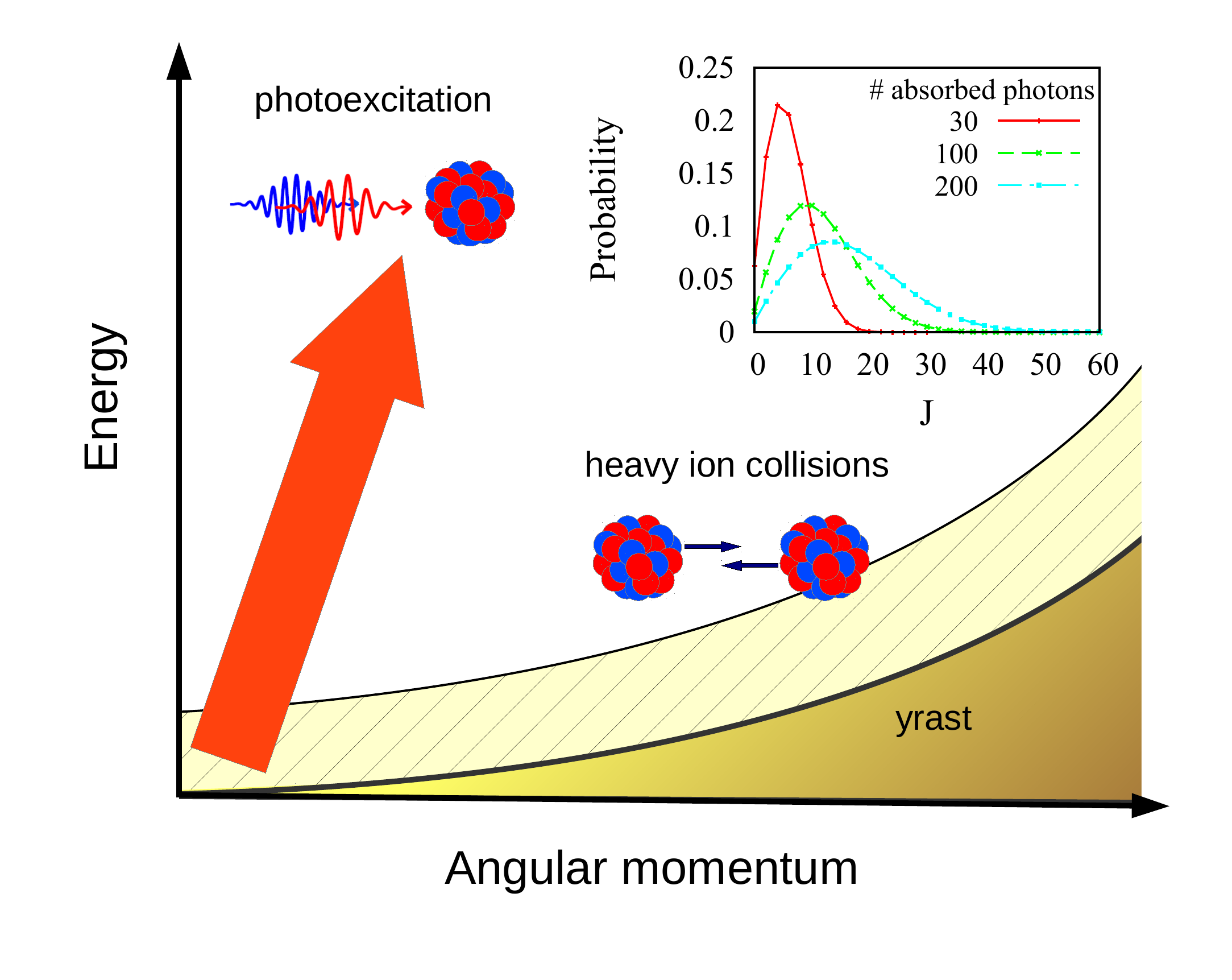}
  \caption{\label{fig1} (Color online) Qualitative illustration of two
    regimes of nuclear excitation. The yrast line defines the minimum 
    energy of a nuclear state with a certain angular momentum. Heavy-ion 
    collisions preferentially excite states close to the yrast line (region depicted by
    hatched area).  Multiple absorption of coherent multi-MeV dipole
    photons involves small transfer of angular
    momentum and leads to compound states several hundred MeV above
    yrast (red arrow). The inset shows the angular-momentum distributions for
    $N_0=30, 100$ and $200$ absorbed dipole photons.}
\end{figure}


We simplify the description further and assume that between two
subsequent photon absorption processes, nuclear equilibration is
complete. Without that assumption, the process must be described in
terms of a time-dependent master equation. That does not seem
justified at this early stage of  theoretical and experimental
development. Photon absorption at excitation energy $E$ is then
governed by the effective absorption rate of an equilibrated compound
nucleus and given by $(N \Gamma)_{\rm eff}(E) = N \Gamma_{\rm dip}
\ \rho_{\rm acc}(E) / \rho_{\rm acc}(E_g)$. Here $\rho_{\rm acc}(E)$
is the density of accessible states and $E_g$ is the energy of the
ground state. We have used the expression for $\rho_{\rm acc}(E)$
based on the Fermi-gas model given in Ref.~\cite{Pal13b}. Our results show that
$(N \Gamma)_{\rm eff}(E)$ slowly decreases with increasing $E$. This supports
our assumption that the spreading width $\Gamma_{\rm sp}$ does not change
significantly with excitation energy.

The consecutive absorption of $N_0$ photons terminates prior to the
end of the laser pulse whenever $(N \Gamma)_{\rm eff}(E)$ is equal to
the largest one of four widths: The width $(N \Gamma)_{\rm ind}(E)$
for induced dipole emission, the width $\Gamma_n(E)$ for neutron
evaporation, the width $(N \Gamma)_{\rm cnt}(E)$ for induced nucleon
emission, and the width $\Gamma_f$ for induced fission. The
expressions for these four widths involve the density $\rho_A(E)$ of
spin-zero states of the target nucleus with mass number $A$ at
excitation energy $E$, or the density $\rho_{\rm acc}(E)$ of
accessible states. For $\rho_A(E)$ we use the expressions given in
Ref.~\cite{Pal13b}. These are valid for high excitation energies $E$
and for $A \gg 1$ and depend on the density $\rho_1(\ve)$ of bound
single-particle states. We have used two continuous forms for
$\rho_1(\ve)$,
\be \rho^{(1)}_1(\ve) = \frac{2 A}{F^2} \ve \ , \ \rho^{(2)}_1(\ve) =
\frac{3 A}{F^3} \ve^2 \ ,
\label{1}
\ee
with $\rho^{(1)}_1(\ve)$ ($\rho^{(2)}_1(\ve)$) used for mass number $A
= 100$ ($A = 200$, respectively). The range of the single-particle
spectrum is $0 \leq \ve \leq V$ with $V = 45$ MeV while the Fermi
energy $F$ was taken as $F = 37$ MeV. For $A = 100$ ($A = 200$) the
$A$-particle level density $\rho_A(E)$ reaches its maximum at an
excitation energy $E_{\rm max} = (2/3) A (V - F) = 533$ MeV (at
$E_{\rm max} = (3/4) A (V - F) = 1200$ MeV, respectively). The
expressions~(\ref{1}) for $\rho_1(\ve)$ were also used~\cite{Pal13b}
to calculate $\rho_{\rm acc}(E)$ and the density of accessible
continuum states $\rho_{\rm cont}(E)$ introduced below.

{\it Induced Dipole Emission}. Probability conservation in the master
equation implies $(N \Gamma)_{\rm ind}(E) = (N \Gamma)_{\rm eff}(E)
\ \rho_A(E - E_L) / \rho_A(E)$. The ratio $\rho_A(E - E_L) / \rho_A(E)$
is very small at excitation energies in the $10$ MeV range but
increases steeply with $E$. Absorption and induced emission become
equal at the maximum $E_{\rm max}$ of $\rho_A(E)$. Substantial
excitation of the compound nucleus by dipole absorption beyond $E_{\rm
  max}$ is impossible because induced dipole emission overcompensates
absorption. In the absence of all other decay mechanisms, the nuclear
occupation probability would hover in a set of states with excitation
energies close to $E_{\rm max}$ until the laser pulse terminates.

{\it Neutron Evaporation}. From the Weisskopf formula we have
$\Gamma_n(E) = (2 \pi)^{- 1} \int_{E_g(A-1)}^{E - E_n} {\rm d} E'
\rho_{A - 1}(E') / \rho_A(E)$. Here $E$ is the excitation energy, $E_n
= V - F$ is the binding energy of the last neutron, and $E_g(A-1)$
($\rho_{A - 1}(E)$) is the ground-state energy (the level density,
respectively) of the nucleus with mass number $A - 1$. In accordance
with our semiquantitative approach we have taken all transmission
coefficients in the Weisskopf formula (i.e., the transmission
probabilities into the individual open neutron channels) equal to
unity. To calculate $E_g(A-1)$ and $\rho_{A - 1}(E)$ we have used the
single-particle level densities in Eqs.~(\ref{1}) with $A \to (A -
1)$. Because of its dependence on level densities, $\Gamma_n(E)$ rises
steeply with excitation energy. The point of intersection with the
curve for $(N \Gamma)_{\rm eff}(E)$ defines the neutron-evaporation
limit of excitation by dipole absorption.

{\it Induced Nucleon Emission}. The occupation probability of
single-particle states above the Fermi energy increases with
increasing excitation energy. Dipole absorption by nucleons in such
states may lead to direct particle emission into the continuum. In
analogy to the expression for $(N \Gamma)_{\rm eff}(E)$, we have $(N
\Gamma)_{\rm cnt}(E) = N \Gamma_{\rm dip} \ \rho_{\rm cnt}(E) /
\rho_{\rm acc}(E_g)$. As in the case of $\rho_{\rm acc}(E)$, the
Fermi-gas model was used~\cite{Pal13b} to calculate the density of
accessible continuum states $\rho_{\rm cnt}(E)$. While for the
calculation of $\rho_{\rm acc}(E)$ only bound single-particle states
(with energies $\ve < V$) are taken into account, for $\rho_{\rm
  cnt}(E)$ only particle-instable single-particle states with energies
$\ve \geq V$ are used. The density of these states was determined by a
fit to results given in Ref.~\cite{Shl92}. We have not attempted to
determine the ratio of protons to neutrons emitted in the process.
This ratio is expected to depend on the height of the Coulomb barrier.

{\it Induced Fission}. According to the Bohr-Wheeler
formula~\cite{Boh39} modified by friction~\cite{Gra83}, $\Gamma_f$
decreases monotonically with increasing friction constant $\beta$. We
use the maximum value $\Gamma_f = (\hbar \omega_1 / (2 \pi)) \exp \{ -
E_f / T \}$ attained at $\beta = 0$. Here $\omega_1$ is the frequency
of the inverted harmonic oscillator that osculates the fission barrier
at its maximum, $E_f$ is the height of the fission barrier, and $T$ is
the nuclear temperature given by $1 / T = (\rm d / \rm d E) \ln
\rho_A(E)$. Very little is known about the temperature dependence of
$\omega_1$ and $E_f$ and of the Strutinsky shell
corrections~\cite{Bra72}. Therefore, our estimate for $\Gamma_f$ is
more qualitative than for the other widths.

\begin{figure}[b]
\vspace{-0.4cm}
  \includegraphics[width=0.45\textwidth]{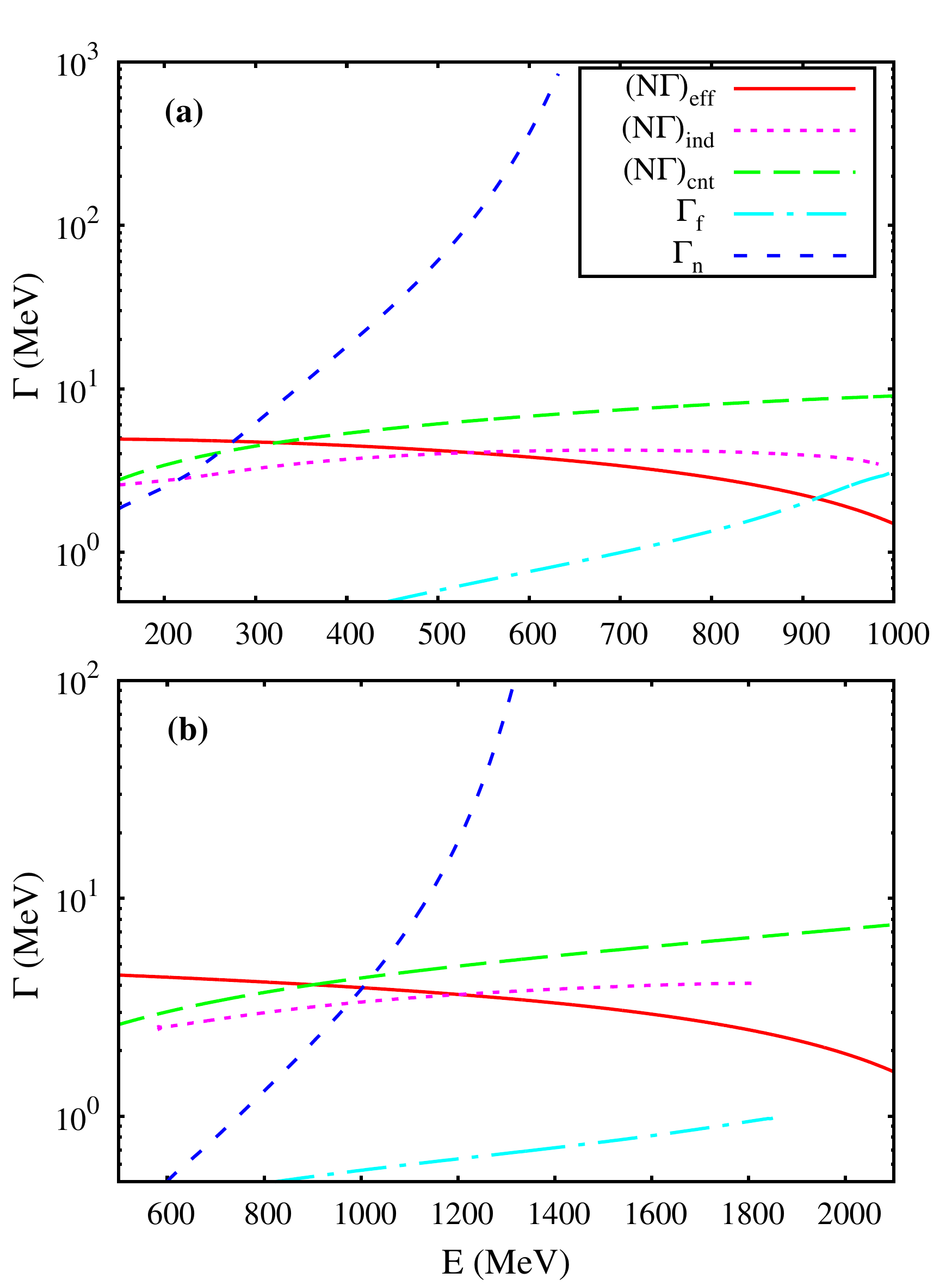}
  \caption{\label{fig2} (Color online) Various widths (in MeV) versus
    excitation energy (in MeV) (a) for $A=100$ and (b) for
    $A=200$. The solid red line depicts $(N \Gamma)_{\rm eff}$, the
    dotted pink line $(N \Gamma)_{\rm ind}$, the long-dashed green
    line $(N \Gamma)_{\rm cnt}$, the dash-dotted light blue line
    $\Gamma_f$, and the short-dashed dark blue line $\Gamma_n$,
    respectively. For $\Gamma_f$ we have used in the calculations
    $\omega_1 = 4$ MeV and $E_f =10$ MeV ($4$ MeV) for $A = 100$ ($A =
    200$, respectively).}

\end{figure}

\begin{figure}[b]
\vspace{-0.4cm}
  \includegraphics[width=0.45\textwidth]{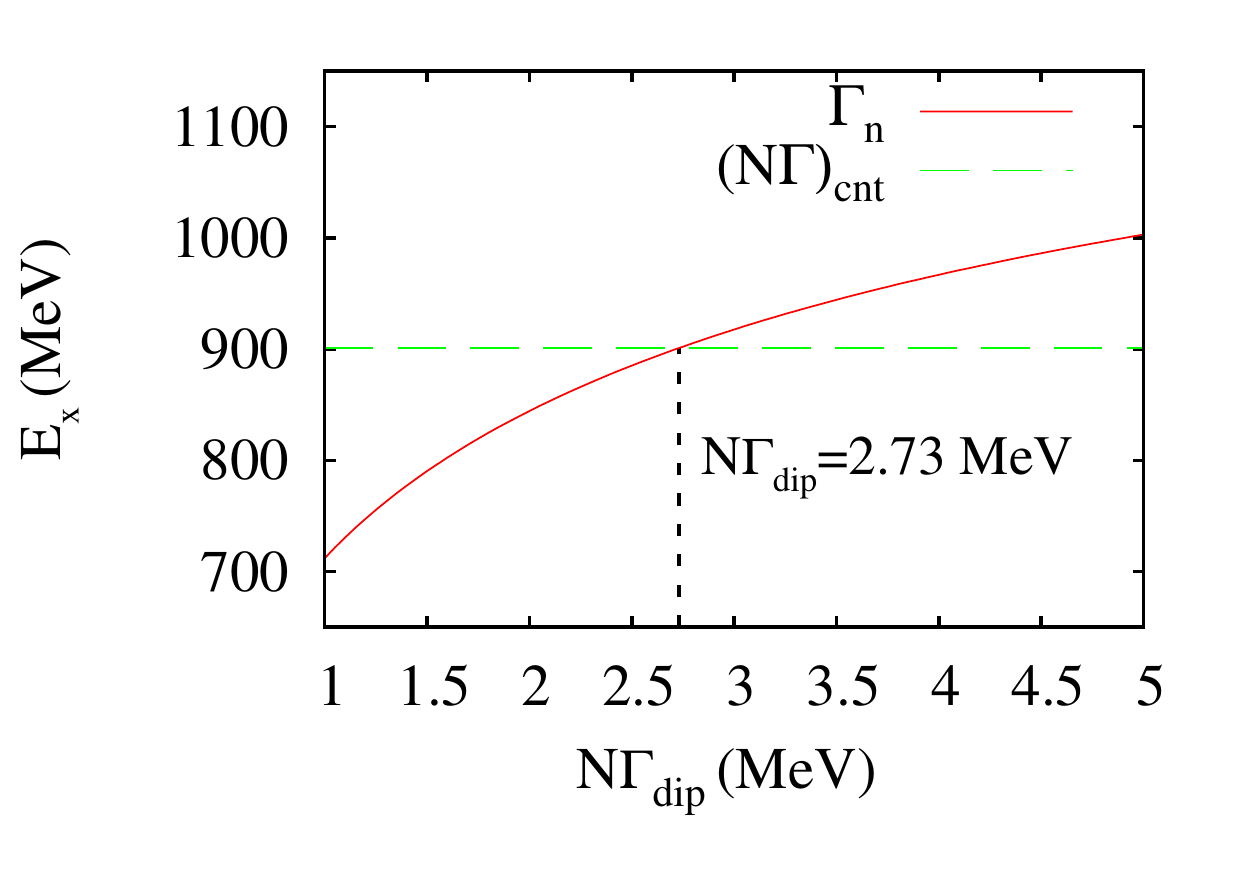}
  \caption{\label{fig3} (Color online) The solid red line (dashed
    green line) gives the energy $E_{\rm x}$ versus $N \Gamma_{\rm dip}$
    where the curves for $(N \Gamma)_{\rm eff}$ and $\Gamma_n$ (for
    $(N \Gamma)_{\rm eff}$ and $(N \Gamma)_{\rm cnt}$, respectively) 
    intersect for A=200.}
\end{figure}

{\it Results}. Fig.~\ref{fig2} shows the five widths (in MeV) for $A =
100$ and $A = 200$ versus excitation energy $E$ (in MeV) and for $N
\Gamma_{\rm dip} = 5$ MeV. We note that $(N \Gamma)_{\rm eff}(E)$
decreases slowly as $E$ increases. At the maxima $E_{\rm max}$ of the
level density $\rho_A(E)$ given above, $(N \Gamma)_{\rm eff}(E)$
intersects with $(N \Gamma)_{\rm ind}(E)$. In all cases considered the
point of intersection of $(N \Gamma)_{\rm eff}(E)$ with either
$\Gamma_n(E)$ or $(N \Gamma)_{\rm cnt}(E)$ lies significantly below
$E_{\rm max}$. Thus, nuclear excitation by dipole absorption is always
limited by neutron evaporation or induced nucleon emission. The
fission width is always smaller than the other widths and does not
terminate dipole absorption, even though we have chosen the
unrealistically large value $\omega_1 = 4$ MeV. For $A = 100$ neutron
evaporation is the dominant process irrespective of the value of $N
\Gamma_ {\rm dip}$. For $A = 200$, however, the competition between
neutron evaporation and induced nucleon emission is decided by $N
\Gamma_ {\rm dip}$. In the case of Fig.~\ref{fig2}(b) ($A = 200$ and
$N \Gamma_ {\rm dip} = 5$ MeV), photon absorption is terminated by
induced nucleon emission. Fig.~\ref{fig3} shows the dependence of the
intersection points of the various widths on the value of $N
\Gamma_{\rm dip}$. While $\Gamma_n$ is independent of $N \Gamma_{\rm
  dip}$, both $(N \Gamma)_{\rm eff}(E)$ and $(N \Gamma)_{\rm cnt}(E)$
depend linearly on $N \Gamma_{\rm dip}$. Therefore, the intersection
point of the two latter curves is fixed at $E_{\rm x} = 901$~MeV. At
$N \Gamma_{\rm dip}=2.73$ MeV, neutron evaporation and induced nucleon
emission exchange their roles in limiting dipole excitation.

If neutron evaporation dominates over induced nucleon emission, a
single highly excited daughter nucleus with mass number $(A - 1)$ is produced. 
 Our values for $\rho_{A - 1}(E)$ show that for $A = 100$ ($A = 200$) the spectrum of evaporated neutrons
falls off (nearly) exponentially with energy  and less than $10$ per
cent of the emitted neutrons have energies in excess of $20$ MeV ($25$
MeV, respectively). Therefore, absorption of only a few photons
suffices to excite the daughter nucleus to energies where another
neutron is emitted. The chain continues. A laser pulse of sufficient
duration thus opens the possibility to reach proton-rich nuclei far
off the line of stability.

Our neglect of proton evaporation may not be justified for nuclei where
proton binding energies are small, especially for proton-rich nuclei.
Moreover, protons may be emitted in significant numbers when induced
emission of protons and neutrons dominates over neutron evaporation.
At the end of the laser pulse we expect a fixed number of nucleons to
be emitted. The resulting final-product nuclei are in highly excited states
and have fixed mass number but a distribution in proton numbers that
ranges from the valley of stability to a very proton-rich nucleus.
The exact details depend on the competition between neutron and proton emission
and are beyond the scope of this paper.

{\it Discussion and Conclusions.} We have studied theoretically the
interaction of a strong coherent zeptosecond laser pulse with
medium-weight and heavy nuclei. The comparison with atom-laser
reactions shows that in the nuclear case the interaction is
comparatively weak so that we always deal with multiphoton
excitations. A novel aspect of the nuclear case is the important role
played by the residual interaction which drives the nucleus towards
statistical equilibrium. Combined with the fact that dipole absorption
dominates all other multipoles, this leads in the quasiadiabatic case
to compound nucleus excitation energies far above yrast.

The main uncertainty in our calculations is due to the various level
densities that determine the five widths. However, each width actually
depends on a ratio of many-body level densities taken at nearly the
same energies and/or mass numbers. Such ratios are much less sensitive
to details of the single-particle level density $\rho_1(\ve)$ in
Eqs.~(\ref{1}) than the many-body level densities themselves.
Therefore, we expect our results not to change drastically when other
values for $\rho_1(\ve)$ are used. Such values could be obtained, for
instance, from a temperature-dependent Hartree-Fock calculation.
Nevertheless, it would be unreasonable to expect that our results
define the critical energies more precisely than to within several
$10$ MeV. Within such errors it seems reasonably safe to say that
neutron evaporation and induced nucleon emission (and not induced
dipole emission) terminate photon absorption, and that nuclear fission
is irrelevant (except perhaps for the heaviest nuclei not considered
here). The competition between neutron evaporation and induced nucleon
emission is so narrow for $A = 200$, however, that either of these
processes may dominate.

Typical maximum excitation energies depend on the intensity and
duration of the laser pulse. With $E_L$ the mean energy per photon,
$N_0 = E / E_L$ photons must be absorbed to reach the high excitation
energies $E$ of up to $\approx 1000$ MeV shown in Figs.~\ref{fig2} and
\ref{fig3}. Even larger values of $N_0$ are needed to reach nuclei far
off the line of stability. The time for the total absorption process
is roughly $N_0 \hbar / (N \Gamma_{\rm dip})$, and the laser pulse
duration must then obey $\hbar / \sigma \geq N_0 \hbar / (N
\Gamma_{\rm dip})$. Hence, large values of $N$, values of $E_L$ in the
$5$ MeV range, and values of $\sigma$ in the $10$ keV range or below
are desirable to exploit the full potential of the process and reach the 
region far above yrast. If the laser pulse lasts long enough,
nuclei far off the line of stability are produced. Then, laser-induced
nuclear reactions promise insight into the structure of proton-rich
nuclei.


\end{document}